\documentclass[journal,onecolumn,10pt]{WSPLC_class}

\usepackage[left=2cm,top=1.78cm,right=2cm]{geometry}
\usepackage{graphicx}
\usepackage{subfig}
\usepackage{amsmath}
\usepackage{epstopdf}
\begin{document}

\title{Novel Grid Topology Estimation Technique Exploiting PLC Modems}

\author{
	\large
	 Federico Passerini and Andrea M. Tonello\\\vspace{6pt}
	\normalsize
	 	Alpen-Adria-Universit\"at Klagenfurt, Klagenfurt, Austria \\ \{federico.passerini, andrea.tonello\}@aau.at
}

\markboth{\hspace{0pt}Tenth Workshop on Power Line Communications, 10-11 October 2016, Paris, France}{}


\maketitle

\vspace{-10pt}
%

\IEEEpeerreviewmaketitle

\section{Introduction}

\IEEEPARstart{P}{ower} Line Communications (PLC) have been extensively deployed both for in-home high speed networking and for smart grid operation management \cite{lampe2016power,7467440}. In this latter context, PLC are used as a mean of controlling and monitoring the power flow of the grid. They enable the exange of information among devices as meters as well as monitoring/control devices and switches. Thus, PLC also foster the developement of routing strategies for both power and data information, as well as coordination algorithms for distributed computation. With these capabilities, the smart grid is able to self control the power flows from different distributed sources to optimize the user power demand, avoid energy waste, and to balance the network loads to obtain the maximum power transfer efficiency.

For all the aforementioned tasks, a fundamental requirement is the knowledge of the network topology, which might not be complete or might not properly track the topological changes due to, for instance, the commutation of power switches. For these reasons, some recent proposals aimed at devising a method to infer the power line distribution network topology in a real-time fashion, without using prior topological information \cite{lampe2013tomography,erseghe2013topology}. These methods propose a two step approach: firstly, the distance between each PLC node and all the others is estimated; then, an algorithm is applied to infer the topology. Although \cite{lampe2013tomography,erseghe2013topology} present different inference algorithms, they both require an estimate of the distance between the PLC modems, which is obtained from estimation of the propagation time of the signal transmitted from one PLC modem to another. 

In this work, we present a novel method to derive the topology of a distribution network that exploits the capability of PLC modems to measure the network admittance, and we report the most significant results.

\section{Approach and results}
PLC modems enable the network admittance measurement at each node of the network. From the knowledge of the network admittance, it is possible to infer the network topology by properly manipulating the equations of the transmission line theory. Differently from \cite{lampe2013tomography,erseghe2013topology}, both the distances between the PLC nodes and the topological information are learned in a single-step algorithm, which saves operational time and limits the error propagation. In particular, we rely on the theorem demonstrated in \cite{DBLP:journals/corr/PasseriniT16}, which states:

	When the cable parameters are known and the impedance measurements are perfect, the topology (graph and branch lengths) of any line network in which 
	\begin{equation*}
		\max_{i,j \in N} d_{i,j} \le \frac{\lambda}{4},
	\end{equation*}
where $d$ is the branch length and $i,j$ are node indexes, can be exactly derived by means of a recursive use of transmission line equations, with the exploitation of the measured network admittances and the available cable parameters and loads.

Our approach has three important properties: firstly, it needs the admittance measurement to be performed at a single frequency, which allows the choice, at every time instant, of the frequency with the highest SNR; secondly, the topology inference capability is in principle not limited by the dimension of the network but by the maximum admitted line length, which is an inverse function of the frequency adopted to perform the measurement; finally, since the network admittance is measured at the transmitter side, the signal-to-noise (SNR) ratio is much higher than the SNR at the receiver. For example, in the Cenelec A band the injected power is in the order of -15~dBm, while the PLC background noise ranges from -70~dBm to -90~dBm \cite{7037264}, so that the SNR is higher than 55~dB. Moreover, the admittance measurement noise is directly related to the classical voltage noise, so that the admittance-to-noise ratio (ANR) can be greater than the SNR, depending on the transmitter output admittance\cite{DBLP:journals/corr/PasseriniT16}. Hence, with a proper tuning of the transmitter admittance, the minimum ANR can reach values up to 85-90~dB.

\begin{figure}
	\centering
	\subfloat[]
		{\includegraphics[width=0.33\textwidth]{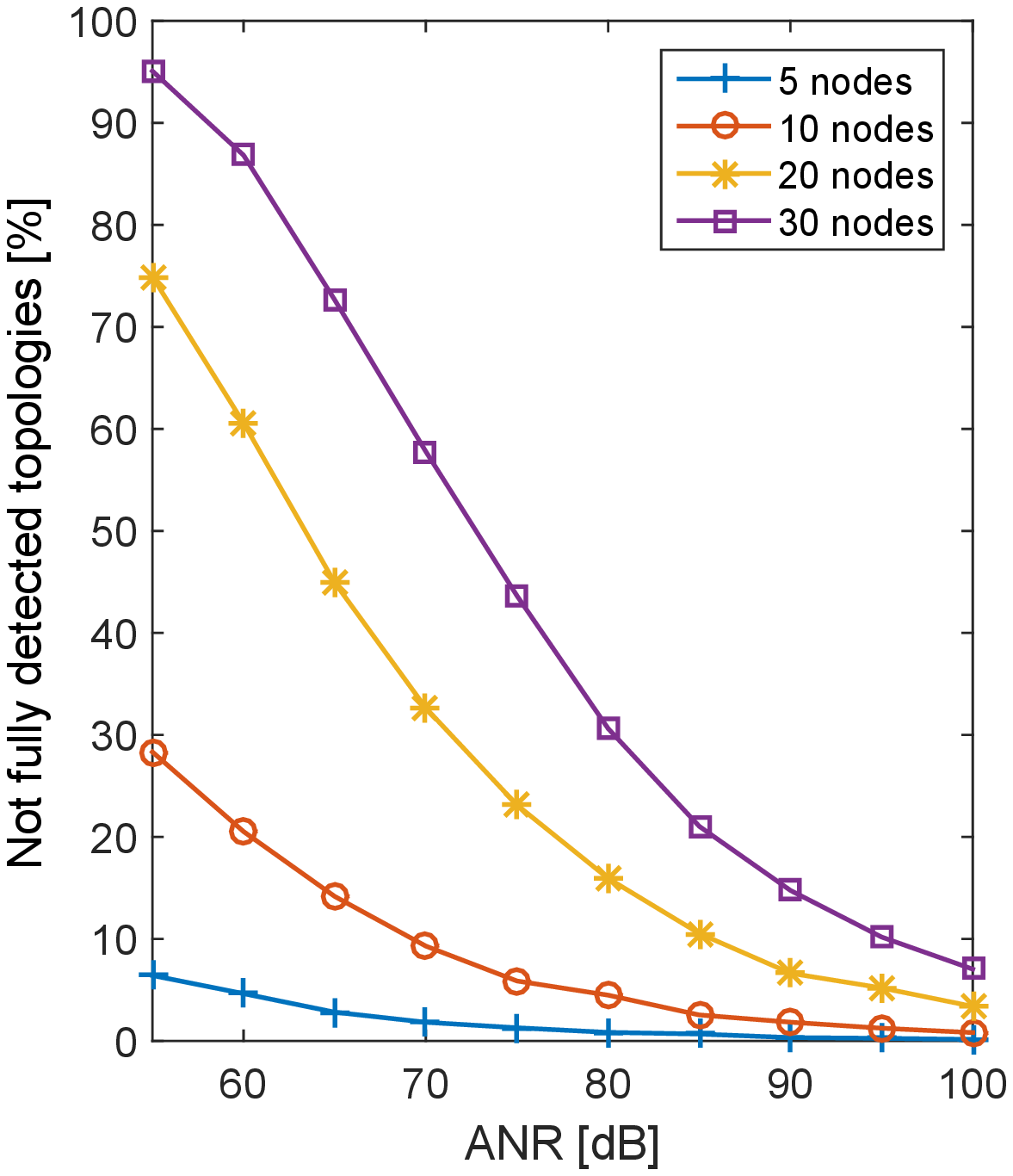}\label{aa}}
	\subfloat[]
		{\includegraphics[width=0.33\textwidth]{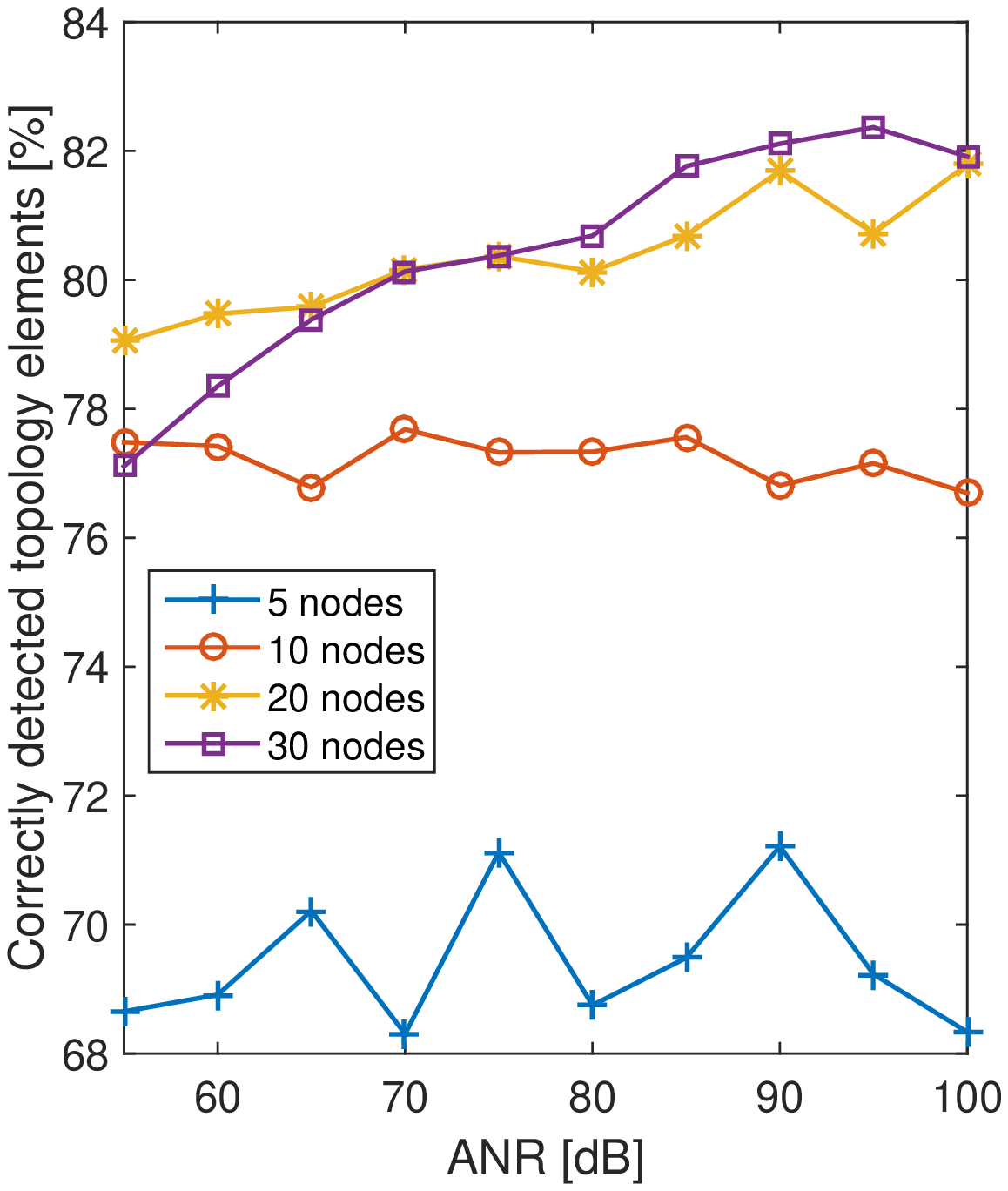}\label{bb}}	
	\caption{Percentage of not fully detected topologies as a function of the SNR (a). Percentage of correctly identified topology elements when the topology is not fully inferred (b). Test parameters: maximum cable length 500~m, measurement frequency 10~kHz.}
	\label{fig:thresholding_effect_all_fit}
\end{figure}

In order to test the above mentioned algorithm, we developed a new full bottom-up PLC network simulator similar to \cite{tonello2011bottomup,tonello2010bottomup} that generates random topologies where the number of branches follows a Poisson distribution. The network admittance at all the nodes of the generated network are then computed and perturbed by colored Gaussian noise. The topology learning algorithm is applied on thousands of network realizations in order to validate it. In Figure~\ref{fig:thresholding_effect_all_fit}, the results for networks with 5 to 30 ndes as a function of the ANR are shown. In this example, we set a maximum cable length of 500~m and measure the network admittances corrupted by noise at 10~kHz. At this frequency the typical SNR in PLC networks (considering just the background noise) is around 60~dB, and an optimal ANR is around 95~dB. Figure~\ref{aa} shows that, for example, the topology of a 30 node network is correctly identified in about 90\% of the realizations, with the mentionned ANR. On the other hand, Figure~\ref{bb} shows for the same scenario the percentage of correctly detected topological elements (connection of two modems and branch length) when the total topology has not been correctly identified. Interestingly, a considerable part of the network is correctly identified even for rather low SNRs.

\section{Conclusion}
This paper has introduced a novel approach for topology estimation in the context of distribution networks, exploiting the capability of the PLC modems to measure the network admittance. Future research will focus on the use of advanced estimation techniques to better estimate the topology in the presence of error in the parameters and of more complex sources of noise.

\bibliographystyle{IEEEtran_preprint}
\bibliography{IEEEabrv,biblio}

 \newcommand{\noop}[1]{}
\begin{thebibliography}{1}
\providecommand{\url}[1]{#1}
\csname url@samestyle\endcsname
\providecommand{\newblock}{\relax}
\providecommand{\bibinfo}[2]{#2}
\providecommand{\BIBentrySTDinterwordspacing}{\spaceskip=0pt\relax}
\providecommand{\BIBentryALTinterwordstretchfactor}{4}
\providecommand{\BIBentryALTinterwordspacing}{\spaceskip=\fontdimen2\font plus
\BIBentryALTinterwordstretchfactor\fontdimen3\font minus
  \fontdimen4\font\relax}
\providecommand{\BIBforeignlanguage}[2]{{%
\expandafter\ifx\csname l@#1\endcsname\relax
\typeout{** WARNING: IEEEtran.bst: No hyphenation pattern has been}%
\typeout{** loaded for the language `#1'. Using the pattern for}%
\typeout{** the default language instead.}%
\else
\language=\csname l@#1\endcsname
\fi
#2}}
\providecommand{\BIBdecl}{\relax}
\BIBdecl

\bibitem{lampe2016power}
L.~Lampe, A.~M. Tonello, and T.~G. Swart, Eds., \emph{Power Line
  Communications: Principles, Standards and Applications from Multimedia to
  Smart Grid}.\hskip 1em plus 0.5em minus 0.4em\relax Wiley, 2016.

\bibitem{7467440}
C.~Cano, A.~Pittolo, D.~Malone, L.~Lampe, A.~M. Tonello, and A.~G. Dabak,
  ``State of the art in power line communications: From the applications to the
  medium,'' \emph{IEEE Journal on Selected Areas in Communications}, vol.~34,
  no.~7, pp. 1935--1952, July 2016.

\bibitem{lampe2013tomography}
M.~O. Ahmed and L.~Lampe, ``Power line communications for low-voltage power
  grid tomography,'' \emph{IEEE Transactions on Communications}, vol.~61,
  no.~12, pp. 5163--5175, Dec. 2013.

\bibitem{erseghe2013topology}
T.~Erseghe, S.~Tomasin, and A.~Vigato, ``Topology estimation for smart micro
  grids via powerline communications,'' \emph{Signal Processing, IEEE
  Transactions on}, vol.~61, no.~13, pp. 3368--3377, July 2013.

\bibitem{DBLP:journals/corr/PasseriniT16}
\BIBentryALTinterwordspacing
F.~Passerini and A.~M. Tonello, ``On the exploitation of admittance
  measurements for wired network topology derivation,'' \emph{IEEE Transactions
  on Instrumentation and Measurements}, \noop{3001}in press. [Online]. Preprint
  available: \url{http://arxiv.org/abs/1607.06417}
\BIBentrySTDinterwordspacing

\bibitem{7037264}
M.~Girotto and A.~M. Tonello, ``Improved spectrum agility in narrow-band {PLC}
  with cyclic block {FMT} modulation,'' in \emph{2014 IEEE Global
  Communications Conference}, Dec 2014, pp. 2995--3000.

\bibitem{tonello2011bottomup}
A.~Tonello and F.~Versolatto, ``Bottom-up statistical {PLC} channel modelling -
  part \uppercase{I}: Random topology model and efficient transfer function
  computation,'' \emph{IEEE Transactions on Power Delivery}, vol.~26, no.~2,
  pp. 891--898, Apr. 2011.

\bibitem{tonello2010bottomup}
------, ``Bottom-up statistical {PLC} channel modelling - part {II}: Inferring
  the statistics,'' \emph{IEEE Transactions on Power Delivery}, vol.~25, no.~4,
  pp. 2356--2363, Oct. 2010.

\end{thebibliography}

\end{document}